\documentclass[12pt,a4paper,twoside,english]{article}

\usepackage[all]{xy}
\CompileMatrices
\xyoption{2cell}
\usepackage{amsmath}
\usepackage{amssymb,latexsym}
\usepackage{graphics}
\usepackage{textcomp}

\def\e{\varepsilon}
\def\bc{\begin{center}}
\def\ec{\end{center}}
\def\be{\begin{equation}}
\def\ee{\end{equation}}
\def\ba{\begin{array}}
\def\ea{\end{array}}
\def\bea{\begin{eqnarray}}
\def\eea{\end{eqnarray}}

\def\lra{\longrightarrow}

\def\xla{\xleftarrow}

\def\bc{\begin{center}}
\def\ec{\end{center}}
\def\ni{\noindent}
\def\AG{\mathrm{Aut}\,(G)}
\def\LG{\mathfrak{g}}
\def\LAG{\partial (\mathfrak{g}) }
\def\g{\gamma}
\def\f{\phi}
\def\F{\Phi}

\def\P{\Pi_1(X)}
\def\PP{\Pi_2(X)}
\def\PX{{\mathcal P}(X)}

\def\B{\mathcal{B}}
\def\C{\mathcal{C}}

\def\O{\mathrm{Ob}}
\def\T{\mathcal{T}}
\def\TB{\mathcal{T}_L (\mathcal{B})}
\def\pic{\xymatrix}
\def\vs{\vspace{5mm}}
\def\nl{\newline}

\begin{document}


\begin{titlepage}

\today\hfill PAR-LPTHE 02/12
\vskip 4cm

\begin{center}
{\bf \huge{Combinatorics of Non-Abelian Gerbes \\ with Connection and Curvature}}
\end{center}

\vskip 5mm
\centerline{Romain Attal \footnote[1]{E-mail: attal@lpthe.jussieu.fr}}
\vskip 5mm
\centerline{
{\em Laboratoire de Physique Th\'{e}orique et Hautes Energies}
\footnote[2]{LPTHE tour 16\,/\,1$^{er}\!$ \'{e}tage,
Universit\'e P. et M. Curie, BP 126 \\ \hspace{1cm} 4 place Jussieu,
F-75252 Paris Cedex 05 (France)}
}
\centerline{\em Universit\'e Pierre et Marie Curie (Paris 6)}

\vs
\vs

{\bf Abstract:} 
We give a functorial definition of $G$-gerbes over a simplicial complex when 
the local symmetry group $G$ is non-Abelian. These combinatorial gerbes are
naturally endowed with a connective structure and a curving. This allows us to 
define a fibered category equipped with a functorial connection over the space 
of edge-paths. By computing the curvature of the latter on the faces of an 
infinitesimal 4-simplex, we recover the cocycle identities satisfied by the curvature
of this gerbe. The link with $BF$-theories suggests that gerbes provide a 
framework adapted to the geometric formulation of strongly coupled gauge theories.

\vspace{1cm}


\end{titlepage}


\section{Introduction}

Fiber bundles with connection are the standard mathematical 
arena for the gauge theories of condensed matter and particle 
physics. Indeed, pointlike excitations carry information along paths in 
space-time and this process is well described by the operation 
of parallel transport. However, in a strongly coupled regime, the 
dynamics of non-Abelian gauge fields allows the appearance of 
extended objects such as loops or membranes and we need a new 
geometrical setting to describe the transport of information on 
surfaces and higher dimensional manifolds. It is known that the 
correct extension of parallel transport to surfaces should be 
defined in gerbes (\cite{Gir}, \cite{BM2}, \cite{Bry}, \cite{MP}) but the 
theory is still in development.

In order to develop our intuition of gerbes, we restrict our approach 
to a discrete setting and use a simplicial complex instead of the 
$\check{\mathrm{C}}$ech-de Rahm complex of a covering of a 
smooth manifold. We extract differential information by considering 
infinitesimal simplexes. Thus we set a simplicial complex $X$ and a Lie group 
$G$ which is our local symmetry group. (In the discrete topology, we can 
use any topological group, but when we consider infinitesimal simplexes, 
we will need a smooth symmetry group). The word "bundle" will always mean 
"combinatorial $G$-bundle with connection" and "gerbes" will in fact mean 
"combinatorial 
$G$-gerbes with 1-connection (connective structure) and 2-connection (curving)". 
We will use informally infinitesimal simplexes and combinatorial differential 
forms \cite{BM2} but without diving into non-standard analysis \cite{Rob} or synthetic 
differential geometry (\cite{K1}, \cite{K2}), where we can handle 
consistently infinitesimal quantities in a broader context than set theory.

\section{Categorical preliminaries}

\subsection{Basic concepts}

We recall here some basic definitions \cite{cwm} in order to set our notations.
A category ${\bf C}$ is defined by two kinds of data : a class 
of objects, $\O\ ({\bf C})$, and, for any pair of objects $x$ and $y$, a set 
of arrows ${\bf C}_{xy}$ (or ${\bf C} (x,y))$. 
These sets are equipped with associative composition laws 
${\bf C}_{xy} \times {\bf C}_{yz} \to {\bf C}_{xz}$ and identities 
$1_x \in {\bf C}_{xx}$, so that the sets ${\bf C}_{xx}$ are monoids. 
When $\O\ ({\bf C})$ is a set, ${\bf C}$ is called a small category. 
For any category ${\bf C}$, its opposite category, ${\bf C}^{\mathrm{op}}$, 
has the same class of objects but the arrows are reversed.
A groupoid is a category in which all the arrows are invertible. The 
monoids ${\bf C}_{xx}$ are then groups. 

A weak 2-category $\C$ is defined by three kinds of data : a class 
of objects, $\O\ (\C)$, and, for any pair of objects $X$ and $Y$, a category 
of 1-arrows $\C_{XY}$. The arrows of $\C_{XY}$ are called the 2-arrows of 
$\C$. We have maps $s_0, t_0, s_1$ and $t_1$, called, respectively, 
0-source, 0-target, 1-source and 1-target, and defined as follows. 
If $X, Y \in \O\ (\C) , u, v \in \O\ (\C_{XY})$ and $\alpha : u \to 
v$, then $s_0 (u) = s_0(v) \equiv X , t_0 (u) = t_0(v) \equiv Y , s_1 
(\alpha) \equiv u$ and $t_1 (\alpha) \equiv v$. 
The composition of these maps thus satisfy the identities
$ s_0 \circ s_1 = s_0 \circ t_1 $ and 
$ t_0 \circ s_1 = t_0 \circ t_1 $
The composition of 1-arrows in $\C$ is simply their 
concatenation, and denoted by $\times$. 
The 2-arrows can be composed either horizontally, if the 
0-target of the first one coincides with the 0-source of the 
second one, or vertically, if the 1-target of the first one 
coincides with the 1-source of the second one. 
When all the 1-arrows and 2-arrows of a 2-category are 
invertible, it is called a 2-groupoid. The invertibility of a 
1-arrow $u:X\to Y$ asserts the existence 
of another 1-arrow $v:Y\to X$ and of invertible 2-arrows 
$\alpha : uv \to 1_X $ and $\beta : vu \to 1_Y $.
$u$ and $v$ are said to be quasi-inverses of each other. 

A category ${\bf C}$ is said monoidal when it is endowed with a bifunctor
${\bf C} \times {\bf C} \to {\bf C} $, with associativity isomorphisms
$a_{x,y,z} : x(yz) \to (xy)z$ satisfying the pentagon axiom

$$a_{x,y,zw}a_{xy,z,w} = (1_x\times 
a_{y,z,w})a_{x,yz,w}(a_{x,y,z}\times 1_w)$$
with an identity object $I$ and with left and right identity isomorphisms
$L_x : Ix\to x$ and $R_x : xI\to x$ such that 

$$a_{x,I,y}(1_x \times L_x)=(R_x \times 1_y)$$
for any $x,y \in \O \ (\bf{C})$. If, moreover, for any $x$ of ${\bf C}$ 
there exists an object ${x}^{*}$ and invertible arrows $\e_x : x {x}^{*}  \to I$ 
and $\eta_x : {x}^{*} x \to I$ then the monoidal category ${\bf C}$ 
is termed compact. A compact, monoidal groupoid is called a gr-category \cite{B2}.

\subsection{Torsors and bitorsors}

Let $G$ be a fixed Lie group. The topological spaces 
endowed with a left (resp. right), continuous, free and 
transitive action of $G$ are called left (resp. right) $G$-torsors. 
$\ell , r , \lambda , \rho , \cdots$ will denote generically such 
actions. A morphism $f : (\ell , X ) \to (\ell ' , X' )$ between two left
$G$-torsors is a continuous $G$-equivariant map : $\ell_g f = f {\ell '}_g$. 
These morphisms are in fact homeomorphisms so that the left (resp. right) 
$G$-torsors and their morphisms form a groupoid $\T_L(G)$ (resp. $\T_R(G)$).
With $\T_L(G)$ and $\T_R(G)$, we can build a group-like category 
$\B$ in the following way \cite{B1}. If we want to define a product of 
two torsors, a one sided action of $G$ isn't enough. However, the left 
action on $(\ell , Y ) \in \O \ (\T_L(G))$ can be contracted with the right action 
on $(X, r) \in \O \ (\T_R(G))$ to give a space $X \times_{G} Y$ (contracted 
product) defined by

\begin{eqnarray*}
X \times_{G} Y & = & X \times Y / \sim \\
(x , \ell_g (y) ) & \sim & (r_g (x) , y) 
\end{eqnarray*}
For $X \times_{G} Y$ to be in the same category as $X$ and $Y$, 
they must all carry a left action and a right action of $G$. A manifold
endowed with such actions is called a $G$-bitorsor. A morphism 
$f : (\ell , X , r) \to (\ell ' , X', r')$ between two bitorsors is a 
continuous map which commutes with both actions on $X$ and $X'$. 
Since $\ell_g (x)$, the image of $x \in X$ under the left action of 
$g \in G$, can be reached by the right action of a unique $h \in G$, 
the correspondance $g \to h=\varphi (g)$ defines a map $\varphi : G \to G$, 
which is actually an automorphism of $G$. Conversely, any $\varphi \in \AG$ 
determines a unique bitorsor from a left $G$-torsor $(\ell , X )$, by 
defining the right action of $g \in G$ as follows :

$$r_g (x)  =  \ell_{\varphi^{-1}(g)} (x) 
\hspace{15mm}\forall \  x \in X , \ \forall \  g \in G $$
$\B $ will denote the gr-category of $G$-bitorsors and $\TB$ 
the weak 2-groupoid of $\B$-torsors. A $\B$-torsor is a category $\C $ 
on which $\B $ acts by equivalences, that is to say each $G$-bitorsor $\beta$ is 
mapped to a functor $F_{\beta}:\C  \to \C $ which admits a quasi-inverse. 
The simplest $\B$-torsors are $\T_L (G)$ and $\T_R (G)$, where the action of 
$(\lambda , \beta , \rho ) \in \B$ on $(\ell , T ) \in \T_L (G)$ twists 
$\ell$ to give the left $G$-torsor $B \times_{G} T \simeq (\lambda , T )$.
The importance of bitorsors in non-Abelian cohomology comes from the fact 
that they compare different torsors and can thus act homogeneously as 
automorphisms on categories which are equivalent to $\T_L (G)$ or $\T_R (G)$.

\section{Bundles as representations of a groupoid of paths}

\subsection{Combinatorial Bundles}

If $X$ is a simplicial complex, we can define the groupoid 
$\P$ as the category whose objects are its vertices and whose arrows are 
its edge-paths, i.e. families $(x_0, x_1, \cdots ,x_n)$ of successive neighbour 
vertices ($n \geq 1$). We identify neither the paths describing the same 
curve at different speeds, nor the paths which differ by "backtrackings" 
(called "thin homotopies" in \cite{MP}). In other words, if $y \neq x$ then 

\begin{eqnarray*}
(\cdots , x , y , y , z , \cdots ) & \neq & (\cdots , x , y , z , \cdots ) \\
(\cdots , x , y , x , z , \cdots ) & \neq & (\cdots ,x , z , \cdots )
\end{eqnarray*}
The insertion or deletion of identical successive vertices is a discrete analogue 
of a local diffeomorphism and the invariance of physical quantities under 
backtracking is the "zigzag symmetry" invoked by A. Polyakov \cite{Pol} 
in the search of a string theory adapted to the description of the confined 
phase of Yang-Mills theory. Both symmetries must be represented 
in any quantum theory of loops.

A left bundle with connection over $X$ is defined as a 
presheaf of left $G$-torsors on $\P$, that is to say a contravariant functor 
$f:\P^{\mathrm{op}} \to \T_L(G)$ which represents the groupoid of paths into the 
groupoid 
of left $G$-torsors. The fibers are the objects $f_x, f_y, f_z, \cdots$ 
associated to the vertices $x, y, z, \cdots$. The connection is determined by 
the images of the edges, $f_{xy} : f_y \to f_x$, which are morphisms 
of left $G$-torsors. Its curvature, $c$ , is the 2-form defined by the holonomies 
along elementary loops :

\begin{eqnarray*}
c &  =  & \sum_{[xyzx]} [xyzx] c_{xyzx}\\
c_{xyzx} & =  & f_{xy}f_{yz}f_{zx}:f_x \to f_x 
\end{eqnarray*}

\ni Each $c_{xyzx}$ is an automorphism of $G$-torsor, i.e. an 
element of $G$, and $c$ is a $G$-valued 2-form, since 
$c_{xzyx} = c_{xyzx}^{-1}$, and $c_{xyzx} =1 \in G$ when
$[xyzx]$ is degenerate.

Two bundles $f , f'$ are decreted to be isomorphic \textit{iff.} there 
exists an invertible natural transformation $g : f' \to f$. In components, 
we have an arrow $g_x : {f'}_x \to f_x$ in $\T_L(G)$ for each $x$, 
such that $f_{xy}g_y = g_x {f'}_{xy}$, i.e. the following square 
commutes : 

\vspace{5mm}
\hspace{6cm}
\pic{
f_x & & f_y \ar[ll]_{f_{xy}} \\
& & \\
{f'}_x \ar[uu]^{g_x} & & {f'}_y \ar[ll]^{{f'}_{xy}} \ar[uu]_{g_y}
}
\vspace{5mm}

\ni $f$ and $f'$ have therefore the same curvature. When $f = f'$, 
$g_x$ and $g_y$ are elements of $G$ and we recover the usual 
expression of gauge transformations.

\subsection{The Bianchi Identity}

We recall here the homotopical interpretation of 
the Bianchi identity satisfied by the connection 1-form, $A$, and its 
curvature 2-form, $F$, in order to extend it to the case of gerbes.
Let $w, x, y$ and $z$ be four pairwise neighbour points, 
i.e. an infinitesimal tetrahedron, with $w$ as base point.  The 
homotopy $(wxyw)(wyzw)(wzxw) \sim (wx)(xyzx)(xw)$, 
which takes place in the 1-skeleton of $X$, implies the 
multiplicative Bianchi identity

\begin{eqnarray}
\boxed{ c_{wxyw} c_{wyzw} c_{wzxw} = f_{wx} c_{xyzx} f_{xw} }
\end{eqnarray}

\ni If we choose an origin in each fiber, $f_{xy}$ is mapped to an element 
of $G$ and, near the identity, we can write $f_{xy}=1+A_{xy}$, $A$ 
being a $\LG$-valued 1-form, i.e. an antisymmetric map 
($A_{yx}=-A_{xy}$) from the set of oriented edges of $X$ to $\LG$. 
We can also write this 1-form as a linear combination of the oriented 
edges, $A=\sum_{(xy)}(xy)A_{xy}$.
The curvature of $A$ is the $\LG$-valued 2-form 
$F=\sum_{[xyzx]}[xyzx]F_{xyzx}$ defined by 

\begin{eqnarray*}
F_{xyzx} & = & c_{xyzx} - 1 \\
&=& (1+A_{xy})(1+A_{yz})(1+A_{zx})-1 \\
&=& (A_{xy} + A_{yz} + A_{zx}) + (A_{xy}A_{yz} + A_{yz}A_{zx}  
+ A_{xy}A_{zx}) \\
F&=& dA+\frac{1}{2} [A,A]
\end{eqnarray*}

\ni Expanding each member of the equation (3) to the second order, we obtain 

\begin{eqnarray*}
c_{wxyw} c_{wyzw} c_{wzxw} &=& 1+F_{wxyw}+F_{wyzw}+F_{wzxw} \\
f_{wx} c_{xyzx} f_{xw} &=& 1+A_{wx}+A_{xw}+F_{xyzx}+
A_{wx}F_{xyzx} + F_{xyzx} A_{xw} \\
&=& 1+F_{xyzx} + [A_{wx},F_{xyzx}] \\
\end{eqnarray*}
and (3) implies the Bianchi identity

\begin{eqnarray*}
{ F_{xyzx}-F_{wxyw}-F_{wyzw}-F_{wzxw}+[A_{wx} , F_{xyzx}] = 0 }
\end{eqnarray*}
usually written in terms of differential forms :
\begin{eqnarray}
\boxed{ dF+[A,F] = 0 }
\end{eqnarray}

\section{Gerbes as representations of a 2-groupoid of surfaces}

\subsection{Combinatorial gerbes}

We know from algebraic topology that the second homotopy \textit{group}
of any space $X$ is always abelian. A better object to study is in fact 
the space of parametrized surfaces. This is a \textit{2-groupoid} and it 
is the basic object in our construction of combinatorial gerbes.
So let's define the 2-groupoid $\PP$ as follows. 
The objects and arrows of $\PP$ are the same as in $\P$ but 
$\PP$ also has 2-arrows generated by the elementary homotopies along 
the oriented 2-cells of $X$. More precisely, two paths with the 
same boundary points, say $\g = (x_0, x_1, \cdots , x_n)$ and 
$\g ' = (x_0, x_1, \cdots , x_p , y , x_{p+1}, \cdots , x_n)$, 
form an elementary 2-arrow of $\PP$ if they differ only by the 
insertion of a vertex $y$ which is a common neighbour to $x_p$ and 
$x_{p+1}$. If the vertex inserted is already present, i.e. if $y=x_p$ or 
$y=x_{p+1}$, then the path does not change and this 2-arrow is considered as 
equal to the identity 2-arrow of this path. These prescriptions define 
the generating 2-arrows of $\PP$. All 2-arrows of $\PP$ are then obtained by 
glueing these generators, either horizontally ($\times$) or vertically 
($\circ$). Thus, a 2-arrow from $\g_0$ to $\g_n$ is a family of paths 
$(\g_0, \cdots , \g_n)$ such that, for all $i\in \{ 0, \cdots , n-1 \}$, 
$(\g_i, \g_{i+1})$ be an elementary 2-arrow of $\PP$. As with paths, 
we don't identify \textit{a priori} the 2-arrows with different 
speeds or with backtrackings. If $\g$ is a neighbour of $\g_i$, then 

\begin{eqnarray*}
(\cdots , \g_{i-1} , \g_i , \g_i , \g_{i+1} ,\cdots) & \neq &
(\cdots , \g_{i-1} , \g_i , \g_{i+1} ,\cdots) \\
(\cdots , \g_{i-1} , \g_i , \g , \g_i , \g_{i+1} ,\cdots) & \neq &
(\cdots , \g_{i-1} , \g_i , \g_{i+1} ,\cdots)
\end{eqnarray*}
The generating 2-arrows of $\PP$ will be written with brackets :

\begin{eqnarray*}
{[xzy]}  =  (xy, xzy) & = & {\mathrm{insertion \ of}\ z\ 
\mathrm{between}\ x\ \mathrm{and}\ y} \\
{[xzy]}^{*}  =  (xzy, xy) & = &  {\mathrm{deletion \ of}\ z\ 
\mathrm{between}\ x\ \mathrm{and}\ y}\\
{[xzyx]}  =  (xx, xzyx) & = & {\mathrm{insertion \ of}\ (zy)\ 
\mathrm{between}\ x\ \mathrm{and}\ x} \\
{[xzyx]}^{*}  =  (xzyx, xx) & = & {\mathrm{deletion \ of}\ 
(zy)\ \mathrm{between}\ x\ \mathrm{and}\ x}
\end{eqnarray*}
We now have the tools to define the combinatorial $G$-gerbes with 
connection as the contravariant 2-functors (prestacks in groupoids) 

$$ \f : \PP^{\mathrm{op}} \to \TB $$

\ni which represent the groupoid of surfaces, $\PP$, in the 
2-groupoid of $\B$-torsors, $\TB$. 
The choice of $\TB$ as a target 2-category 
is a natural one, because its 2-arrows are mapped to elements 
of $G$, once we have chosen an equivalence between its objects 
and the gauge gr-category $\B$. This compact definition hides 
much combinatorial geometry. For example, we will explore in the following 
section the cocycle relations satisfied by the curvature of $\f$. 
Since the fibers are categories $\f_x, \f_y, \cdots$ 
and the edges are mapped to the 1-connection functors $\f_{xy}:\f_y \to 
\f_x $, we can define its (functor valued) curvature 2-form 
$C  =  \sum_{[xyzx]} [xyzx] C_{xyzx} $ by

$$ C_{xyzx}  =   \f_{xy} \f_{yz} \f_{zx} $$

\ni It is important not to confuse $C_{xyzx}$, which is given 
by the usual composition of functors, and the horizontal compostion 
$\f_{xyzx} = \f_{xy} \times \f_{yz} \times \f_{zx}$ which is their 
concatenation, i.e. a fibered category with 1-connection over the path 
$(xyzx)$. The functor $C_{xyzx}$, associated to the 2-cell $[xyzx]$, 
is an equivalence of the $\B$-torsor $\f_x$, therefore it is naturally 
equivalent to the functorial action of some bitorsor $\beta_{xyzx}\in \O\ (\B)$
satisfying 

\begin{eqnarray*}
\beta_{xyzx} \beta_{xzyx} &=& 1_{\B} \\
\f_{xy} \beta_{yzxy} &=& \beta_{xyzx} \f_{xy}
\end{eqnarray*}
We will keep the same notation for this bitorsor, its functorial 
action on $\f_x$, and the corresponding element of $\AG$. 
But $\f$ provides us with natural equivalences which 
represent in $\TB$ the corresponding elementary homotopies. 
More precisely, the 2-arrows

\begin{eqnarray*}
\f_{[xzy]} &:& \f_{xz} \times \f_{zy} \lra \f_{xy} \\
\f_{[xzy]^*} &:& \f_{xy} \lra \f_{xz} \times \f_{zy} \\
\f_{[xyzx]} &:& \f_{xy} \times \f_{yz} \times \f_{zx} \lra \f_{xx} \\
\f_{[xyzx]^*} &:& \f_{xx} \lra \f_{xy} \times \f_{yz} \times \f_{zx}
\end{eqnarray*}
induce natural transformations between the compositions 
of 1-connection functors 

\begin{eqnarray*}
K_{[xzy]} &:& \f_{xz} \f_{zy}\beta_{yzxy} \lra \f_{xy} \\
K_{[xzy]^*} &:& \f_{xy} \lra \f_{xz}\f_{zy} \beta_{yzxy} \\
K_{[xzyx]} &:& C_{xzyx} \beta_{xyzx} \lra 1_{\f_x} \\
K_{[xzyx]^*} &:& 1_{\f_x} \lra C_{xzyx} \beta_{xyzx} 
\end{eqnarray*}
such that, for any $a \in \O (\f_y)$, we have

\begin{eqnarray*}
K_{[\cdots ]^*} &=& (K_{[\cdots ]})^{-1} \\
K_{[xzy]} (a) &=& \f_{xy} ( K_{[yxzy]} (a) ) \\
&=& K_{[xzyx]}(\f_{xy} (a))
\end{eqnarray*}
Similarly, $\f$ induces natural transformations 

\begin{eqnarray*}
{\bar{K}}_{[xzy]} &:& \beta_{xyzx}\f_{xz} \f_{zy} \lra \f_{xy} \\
{\bar{K}}_{[xzy]^*} &:& \f_{xy} \lra \beta_{xyzx}\f_{xz}\f_{zy} \\
{\bar{K}}_{[xzyx]} &:& \beta_{xyzx}C_{xzyx} \lra 1_{\f_x} \\
{\bar{K}}_{[xzyx]^*} &:& 1_{\f_x} \lra \beta_{xyzx}C_{xzyx} 
\end{eqnarray*}
satisfying identical conditions. (Thus it seems possible to break the chiral 
symmetry by giving to the variables $K$ and $\bar{K}$ different dynamics.)
For example, $K_{[xzy]}$ represents the elementary homotopy from 
$(xy)$ to $(xzy)$ as a 2-arrow in the reversed direction ($\f$ being 
contravariant) and whose component at $a$ is the arrow 

$$ K_{[xzy]}(a) : \f_{xz} \f_{zy} \beta_{yzxy} (a) \lra \f_{xy} (a) $$
The 2-functoriality of $\f$ implies the naturality of $K_{[xzy]}$, 
i.e. for each arrow $u : a\to b$ in $\f_y$ we have a commutative diagram

\vspace{5mm}
\hspace{4cm}
\pic{
\f_{xz} \f_{zy} \beta_{yzxy}(a) \ar[dd]_{\f_{xz} \f_{zy} \beta_{yzxy}(u)} \ar[rr]^{\, 
K_{[xzy]}(a)} & &
\f_{xy}(a) \ar[dd]^{\f_{xy}(u)} \\
& & \\
\f_{xz} \f_{zy}\beta_{yzxy}(b) \ar[rr]_{\, K_{[xzy]}(b)} & & \f_{xy}(b) 
}
\vspace{5mm}

\ni i.e. the relation

\begin{eqnarray}
\boxed{ \f_{xy}(u) \circ K_{[xzy]}(a) = K_{[xzy]}(b) \circ \f_{xz} 
\f_{zy}\beta_{yzxy}(u) }
\end{eqnarray}
Analogously, for $[xzyx]$ :
\begin{eqnarray}
\boxed{ u \circ K_{[xzyx]}(a) = K_{[xzyx]}(b) \circ C_{xzyx}\beta_{xyzx}(u) }
\end{eqnarray}
We have to consider the paths with and without backtrackings as 
different because the degenerate 2-cell $[xyx]$ is mapped to a 2-arrow

$$\f_{[xyx]} : \f_{xy} \times \f_{yx} \to \f_{xx}$$
which is a part of the combinatorial  data of the 2-functor $\f$
and amounts to the choice of a quasi-inverse of the 1-connection 
(see \cite{BM1}, \S (7.4)). Indeed, if we choose 
$\beta_{xyyx}  =  1_{\f_x}$, then $ K_{[xyx]}$ is precisely 
the arrow we need to invert $\f_{xy}$, since

$$ K_{[xyx]} : \f_{xy} \f_{yx} \beta_{xyyx} = \f_{xy} \f_{yx} \to 1_{\f_x} $$
A local gauge fixing is obtained by choosing an object $\alpha_x$ in $\f_x$ 
for each vertex $x$ of an infinitesimal tetrahedron. Then, each object $a$ of 
$\f_x$ defines (the isomorphy class of) a bitorsor $b_{xa}$ by

$$b_{xa} (\alpha_x) \simeq a $$
Let $\LAG$ be the Lie algebra of $\AG$, i.e. the set of derivations of $\LG$.
We can define the $\LAG$-valued 
1-form $\mu  =  \sum_{(xy)} (xy) \mu_{xy} $  by

$$ b_{y,  {\f_{xy} (\alpha_x)} }  =  1_{\AG} + \mu_{xy} $$
and the $\LG$-valued 2-form $B =  \sum_{[xyzx]} [xyzx] B_{xyzx} $ by

$$ K_{[xyzx]}  =  1_G + B_{xyzx}$$
Moreover, the expansion of $\beta$ near the identity provides the "fake curvature" 
\cite{BM1}, 
which is the $\LAG$-valued 2-form $\nu  =  \sum_{[xyzx]} [xyzx] \nu_{xyzx}$ defined by

$$ \beta_{xyzx}  =  1_{\AG} + \nu_{xyzx} $$ 
The naturality condition (3 or 4) induces the following identity in $\AG$ :

$$ (\mathrm{Ad}_{1+B_{yxzy}}) (1+\mu_{xy}) = (1+\mu_{xz}) (1+\mu_{zy}) (1+\nu_{yzxy}) $$
Expanding it to the second order, we obtain an identity in $\LAG$ :

\begin{eqnarray*}
{ \mu_{xz}+\mu_{zy}+\mu_{yx}+\mu_{xz}\mu_{zy} = \nu_{yxzy} + [B_{yxzy} , \cdots ] }
\end{eqnarray*}
the limit of which is 

\begin{eqnarray}
\boxed{ \nu = d\mu + \mu^2 - [B , \cdots ] }
\end{eqnarray}
($[X, \cdots ] \in \LAG $ denotes the adjoint action of $X \in \LG$). 
This identity can also be taken as the definition of $\nu$.

\subsection{The fibered categories induced by a gerbe over the space of paths}

In this section, we are going to define two fibered categories over the space of 
paths of $X$, $\F$ and $\bar{\F}$, each one equipped with a functorial 1-connection 
($\F$ and $\bar{\F}$ are related by a reversion of arrows).
From the 2-groupoid $\PP$, we first define the groupoid $\PX$ whose objects 
are the 1-arrows of $\PP$ and whose arrows are the 2-arrows of $\PP$. 
The fiber of $\F$ above $\g = (x_0 , \cdots , x_n)$, 
is defined as the category $\F_\g$ whose objects are the sections of $\f$ above $\g$ :

\begin{eqnarray*}
\O \ (\F_\g) &=& \lbrace (a,u) = (a_0 , u_{01} , a_1 , \cdots , a_{n-1} , u_{n-1,n} , 
a_n)\ \vert \\
& & \quad  a_i \in \O\ (\f_{x_i}) \ \mathrm{and} \ u_{j,j+1} : 
\f_{x_j , x_{j+1}} (a_{j+1}) \to a_j \rbrace \\
\end{eqnarray*}
and whose arrows are the homotopies between such sections :

$$
\F_\g \big( (a,u) ; (b,v) \big) = 
\lbrace (\alpha_i : a_i \to b_i)_{0 \leq i \leq n} \ \vert \
\alpha_i \circ u_{i,i+1} = v_{i,i+1} \circ \f_{x_i , x_{i+1}}(\alpha_{i+1}) \rbrace
$$
The 1- and 2-connection of $\f$ combine to induce a functorial connection on $\F$. 
Indeed, if $(\g , \g ')$ is an oriented edge of $\PX$, and if $\g$ and $\g'$
differ by the 2-cell $[xzy]$, we define $\F_{\g \g '} : \F_{\g '} \to \F_\g$ by the 
following action on the objects of $\F_{\g '}$ : 

\begin{eqnarray*}
& & \F_{\g \g '} (\cdots , a_x , u_{xz} , a_z , u_{zy} , a_y , \cdots) \\
& & \quad =  (\cdots , a_x , u_{xz} \circ \f_{xz} (u_{zy}) \circ K_{[xzy]^*}(a_y) , 
\beta_{yxzy} (a_y) , \beta_{yxzy}(\cdots))
\end{eqnarray*}
where the dots on the left of $a_x$ denote the unchanged entries and 
all the entries on the right of $a_y$ are twisted by $\beta_{yzxy}$ acting on the 
corresponding fiber. 
The functor $\F_{\g'\g} : \F_\g \to \F_{\g'}$ acts as follows :

\begin{eqnarray*}
& & \F_{\g'\g} (\cdots , b_x , v_{xy} , b_y , \cdots) \\
& & \quad =  (\cdots , b_x , v_{xy} \circ K_{[xzy]} (C_{yxzy}\beta_{yzxy}(b_y)) , 
\f_{zy} 
\beta_{yzxy} (b_y) , 
\f_{zy} \beta_{yzxy} (1_{b_y}) , \beta_{yzxy}(b_y) , \beta_{yzxy}(\cdots)) 
\end{eqnarray*}
The initial entries $v_{xy}$ and $ b_y$ are deleted and the entries on the right 
of $b_y$ are twisted by $\beta_{yxzy}$ acting on the corresponding fiber.
One can check easily that $ \F_{\g'\g} $ and $ \F_{\g\g'} $ are quasi-inverses 
one of another. In the definition of the 3-curvature of $\f$, given below, we will also
need the expression of the connection functor $\F_{(xx,xyzx)}$ which maps 
the category of sections of $\f_{xyzx}$ into that of sections of $\f_{xx}$. 
The image of 
$(a_{0x} , u_{xy} , a_y , u_{yz} , a_z , u_{zx} , a_{1x} ) \in \O \ (\F_{xyzx})$ 
via $\F_{(xx , xyzx)}$ is 

\begin{eqnarray*}
& & \F_{(xx , xyzx)} (a_{0x} , u_{xy} , a_y , u_{yz} , a_z , u_{zx} , a_{1x} ) \\
& & \quad =  \big( a_{0x} , u_{xy} \circ \f_{xy}(u_{yz}) \circ \f_{xy} \f_{yz}(u_{zx}) 
\circ K_{[xyzx]^*}(a_{1x}) , \beta_{xyzx} (a_{1x}) \big)
\end{eqnarray*}
Conversely, $\F_{(xyzx , xx)}$ maps $(b_{0x} , v_{xx} , b_{1x})$ to

\begin{eqnarray*}
& & \F_{(xyzx , xx)} (b_{0x} , v_{xx} , b_{1x}) = \\
& & \quad \big( b_{0x} , 
v_{xx} \circ K_{[xyzx]}(b_{1x}) , 
\f_{yz}\f_{zx} \beta_{xzyx} (b_{1x}) , 
\f_{yz}\f_{zx} \beta_{xzyx} (1_{b_{1x}}) , \\
& & \quad \f_{zx} \beta_{xzyx} (b_{1x}) , 
\f_{zx} \beta_{xzyx} (1_{b_{1x}}) , 
\beta_{xzyx} ({b}_{1x}) \big)
\end{eqnarray*}
When we move the path $\g = (x_0 , \cdots , x_n)$ at an end point, 
to the neighbour path $\g' = (x_0 , \cdots , x_{n-1} , y)$, 
$y$ being a neighbour of $x_n$ and of $x_{n-1}$, the sweeping functor 
$\F_{\g' \g}$ acts on $(\cdots , a_{n-1} , u_{n-1} , a_n)$ as follows :

$$ 
\F_{\g' \g} (\cdots , a_{n-1} , u_{n-1} , a_n) = 
\big( \cdots , a_{n-1} , u_{n-1} \circ K_{[x_{n-1} y x_n ]^*} (a_n) , 
\f_{y x_n} \beta_{x_n y x_{n-1} x_n} (a_n) \big) 
$$
The objects of $\bar{\F}$ are defined like those of $\F$ but with reversed arrows : 

\begin{eqnarray*}
\O (\bar{\F}) &=& \lbrace (a,u) = (a_0 , u_{01} , a_1 , \cdots , a_{n-1} , u_{n-1,n} , 
a_n)\ \vert \\
& & \quad  a_i \in \O\ (\f_{x_i}) \, 
\mathrm{and} \,  u_{j,j+1} : a_j \to \f_{x_j x_{j+1}} (a_{j+1}) \rbrace \\
\end{eqnarray*}
The elements of $\bar{\F} ( (a,u) ; (b,v) )$ are defined as homotopies between 
the objects $(a,u)$ and $(b,v)$ and the connection ${\bar{\F}}_{\g \g '}$ is defined by 
formulas similar to those for $\F$ but using $\bar{K} : \beta C \to 1$ and 
$\bar{K}^{-1} : 1 \to \beta C$ instead of $K$ and $K^{-1}$.

We define the 3-curvature of $\f$ as the 3-form $\Omega$ which, 
to each oriented 3-cell $\sigma = \langle vwxyv \rangle$ such that
$ v = s_0 (\sigma) = t_0 (\sigma) $ and $(vv) = s_1 (\sigma) = t_1 (\sigma) $,
associates the 2-arrow $\Omega_{\sigma} : \f_{vv} \to \f_{vv}$ obtained by 
composition of the sweeping functors associated to the four oriented faces of this 
3-cell. 
Let's now compute $\Omega_\sigma$ when the boundary of $\sigma$ 
is represented by the following pasting scheme, swept from top to bottom :

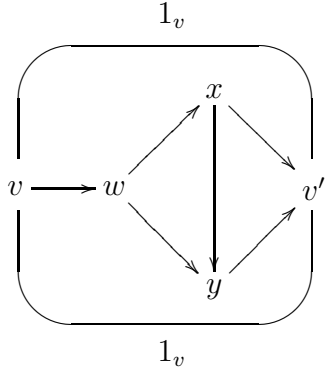
\unitlength=1cm

\vspace{5mm}
\hspace{4cm}
\begin{picture}(4,4)(-1,-3)
\put(2.2,-.9){\oval(3.9,3)[t]}
\put(2.2,-1.6){\oval(3.9,3)[b]}
\put(2.1,.9){$1_v$}
\put(2.1,-3.6){$1_v$}
\pic{
& & x \ar[dd] \ar[dr] & \\
v \ar[r] & w \ar[ur] \ar[dr] & & v' \\
& & y \ar[ur] & 
}
\end{picture}
\vspace{1cm}

\ni Let $(b \xla{\alpha} a) \in \O (\F_{vv'})$ and let's twist $a$ by the 1-arrows 
$C \beta$ associated to the boundaries of the 2-cells swept successively. 
This defines a sequence of objects $a_i \in \O (\f_v)$ 

\begin{eqnarray*}
a_1 &=& C_{vwyv} \beta_{vywv} (a) \\
a_2 &=& \f_{vy} C_{ywxy} \beta_{yxwy} \f_{yv} (a_1) \\
a_3 &=& C_{vxyv} \beta_{vyxv} (a_2) \\
a_4 &=& C_{vwxv} \beta_{vxwv} (a_3)
\end{eqnarray*}
The $K$ arrows corresponding to these 2-cells are 

\begin{eqnarray*}
K_{[vwyv]^*} (a) : a &\lra & a_1 \\
\f_{vy} \big( K_{[ywxy]^*} (\f_{yv} (a_1)) \big) : a_1 & \lra & a_2 \\
K_{[vxyv]} (a_2) : a_2 & \lra & a_3 \\
K_{[vwxv]} (a_3) : a_3 & \lra & a_4
\end{eqnarray*}
The composition of these four arrows defines the 3-curvature arrow $\Omega_\sigma (a)$ :

\begin{eqnarray}
\boxed{ \Omega_\sigma (a)  =  K_{[vwxv]} (a_3) \circ K_{[vxyv]} (a_2) \circ 
\f_{vy} \big( K_{[ywxy]^*} (\f_{yv} (a_1)) \big) \circ K_{[vwyv]^*} (a) }
\end{eqnarray}

\ni Choosing  a local section of $\f$, we can expand the 1-connection as 
$\f_{xy} = 1_\AG +\mu_{xy}$ and the 2-connection as $K = 1_G + B$. 
Since $(1_\B - C)$ is an infinitesimal of second order, the expansion of 
$\Omega_\sigma (a) = 1_G + \omega_{vwxyv}$ to the third order gives 

$$
1 + \omega_{vwxyv} = [1 + B_{vwxv}] [1 + B_{vxyv}] 
\big[ (1 + \mu_{vy}) \cdot (1 - B_{ywxy}) \big] [1 - B_{vwyv}] 
$$
where the dot denotes the action of $\LAG$ on $\LG$. 
Using the antisymmetry of $\omega$ and $B$, we deduce
$$
{\omega_{vywxv} = B_{ywxy} - B_{vywv} - B_{vyxv} - B_{vxwv} + \mu_{vy} \cdot B_{ywxy} }
$$
and the expression of the curvature 3-form $ \omega  =  \sum_{\langle wxyzw \rangle} 
{\langle wxyzw \rangle } \omega_{wxyzw} $ in terms of differential forms becomes 

\begin{eqnarray}
\boxed{ \omega  =  dB+\mu \cdot B }
\end{eqnarray}

\subsection{Cocycle identities}

In the proof of the Bianchi identity, we have used a homotopy 
in the 1-skeleton of an infinitesimal tetrahedron. By analogy, 
in order to obtain the 3-cocycle identities satisfied by $\f$, 
we can use a homotopy in the 2-skeleton of an oriented 4-simplex of $X$. 
So let $v, w, x, y$ and $z$ be five pairwise neighbour vertices of $X$. 
Let's choose $v$ as base point, let's double it into $v$ and $v'$
and let's sweep the boundary of the four tetrahedra containing $v$, according 
to the following pasting schemes, from top to bottom : \newline

First face ($F_1$) :

\hspace{4cm}
\begin{picture}(4,4)(-1,-3)
\put(2.2,-.9){\oval(3.9,3)[t]}
\put(2.2,-1.6){\oval(3.9,3)[b]}
\put(2.1,.9){$1_v$}
\put(2.1,-3.6){$1_v$}
\pic{
& & x \ar[dd] \ar[dr] & \\
v \ar[r] & w \ar[ur] \ar[dr] & & v' \\
& & y \ar[ur] & 
}
\end{picture}
\vspace{5mm}

$$ F_1 = \f_{[vwxv]} \circ (1_{\f_{vwx}} \times \f_{[xyv]}) \circ (1_{\f_{vw}} 
\times \f_{[wxy]^*} \times 1_{\f_{yv}}) \circ \f_{[vwyv]^*} $$

\vspace{5mm}

Second face ($F_2$)  :

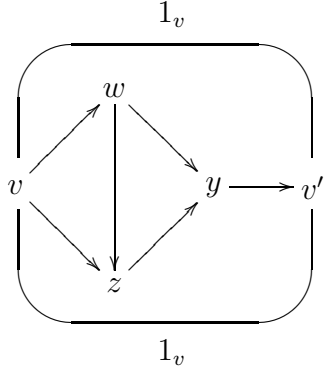
\hspace{4cm}
\begin{picture}(4,4)(-1,-3)
\put(2.2,-.9){\oval(3.9,3)[t]}
\put(2.2,-1.6){\oval(3.9,3)[b]}
\put(2.1,.9){$1_v$}
\put(2.1,-3.6){$1_v$}
\pic{
& w \ar[dr] \ar [dd]& & \\
v \ar[ur] \ar[dr] & & y \ar[r] & v' \\
& z \ar[ur] & &  
}
\end{picture}
\vspace{5mm}

$$ F_2 = \f_{[vwyv]^*} \circ (1_{\f_{vw}} \times \f_{[wzy]} \times 1_{\f_{yv}}) 
\circ (\f_{[vwz]^*} \times 1_{\f_{[zyv]}}) \circ \f_{[vzyv]^*} $$

\vspace{5mm}

Third face ($F_3$)  :

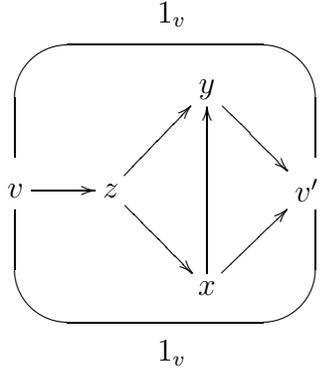
\hspace{4cm}
\begin{picture}(4,4)(-1,-3)
\put(2.2,-.9){\oval(4,3)[t]}
\put(2.2,-1.6){\oval(4,3)[b]}
\put(2.1,.9){$1_v$}
\put(2.1,-3.6){$1_v$}
\pic{
& & y \ar[dr] & \\
v \ar[r] & z \ar[ur] \ar[dr] & & v' \\
& & x \ar[ur] \ar[uu] & 
}
\end{picture}
\vspace{5mm}

$$F_3 = \f_{[vzyv]} \circ (1_{\f_{vz}} \times \f_{[zxy]} \times 1_{\f_{yv}}) 
\circ (1_{\f_{vzx}} \times \f_{[xyv]^*}) \circ \f_{[vzxv]^*} $$

\vspace{5mm}

Fourth face ($F_4$)  :

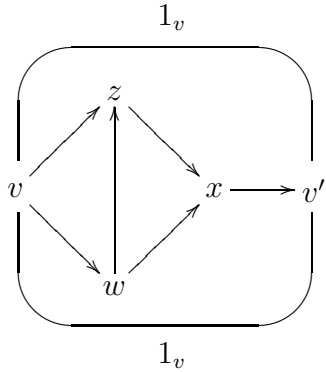
\hspace{4cm}
\begin{picture}(4,4)(-1,-3)
\put(2.2,-.9){\oval(3.9,3)[t]}
\put(2.2,-1.6){\oval(3.9,3)[b]}
\put(2.1,.9){$1_v$}
\put(2.1,-3.6){$1_v$}
\pic{
& z \ar[dr] & & \\
v \ar[ur] \ar[dr] & & x \ar[r] & v' \\
& w \ar[uu] \ar[ur] & & 
}
\end{picture}
\vspace{5mm}

$$ F_4 = \f_{[vzxv]} \circ (\f_{[vwz]} \times 1_{\f_{zxv}}) \circ (1_{\f_{vw}} 
\times \f_{[wzx]^*} \times 1_{\f_{xv}}) \circ \f_{[vwxv]^*} $$

\vspace{5mm}

\ni The composition of these four 2-arrows, which are loops in loop space, 
amounts to sweeping the boundary of the tetrahedron opposite to $v$ 
(fifth face, $F_5$) :

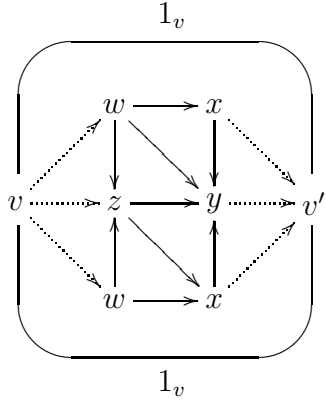
\hspace{4cm}
\begin{picture}(5.5,5.5)(-1,-3.5)
\put(2.2,-.9){\oval(3.9,3.5)[t]}
\put(2.2,-1.6){\oval(3.9,3.5)[b]}
\put(2.1,1.1){$1_v$}
\put(2.1,-3.8){$1_v$}
\pic{
& w \ar[d] \ar[dr] \ar[r] & x \ar[d] \ar@{.>}[dr] & \\
v \ar@{.>}[ur] \ar@{.>}[dr] \ar@{.>}[r] & z \ar[r] \ar[dr] & y \ar@{.>}[r] & v' \\
& w \ar[u] \ar[r] & x \ar[u] \ar@{.>}[ur] & 
}
\end{picture}
\vspace{5mm}

\begin{eqnarray*}
F_5 &=& \f_{[vwxv]} 
\circ ( 1_{\f_{vwx}} \times \f_{[xyv]} ) 
\circ ( 1_{\f_{vw}} \times \f_{[wxy]^*} \times 1_{\f_{yv}} ) 
\circ ( 1_{\f_{vw}} \times \f_{[wzy]} \times 1_{\f_{yv}} ) 
\circ \\
& & (\f_{[vwz]^*} \times  1_{\f_{zyv}} ) 
\circ ( \f_{[vwz]} \times \f_{[zxy]} \times 1_{\f_{yv}} ) 
\circ ( 1_{\f_{vw}} \times \f_{[wzx]^*} \times \f_{[xyv]^*} ) 
\circ \f_{[vwxv]^*} 
\end{eqnarray*}

\vspace{5mm}

\ni The functors defined by these five pasting schemes map $\F_{vv'}$ to itself. 
If we choose a section of $\f$ above this 4-simplex, i.e. an object of reference 
in each fiber, these functors are reduced to the action of an element of $G$ 
on the arrows of $\f_v$ and the identity of these 2-arrows becomes an equality 
in $G$. By expanding the product $F_1 F_2 F_3 F_4$ of the first four faces to the 
fourth order, and equating it to $F_5$, we obtain a combinatorial representation
of the cocycle identity satisfied by $\omega$, which is easier to prove directly with
differential forms :

\begin{eqnarray*}
d \omega + \mu \cdot \omega &=& d (dB +\mu \cdot B) + \mu \cdot (dB +\mu \cdot B) \\
&=& d \mu \cdot B - \mu \cdot dB +\mu \cdot dB +\mu^2 \cdot B \\
&=& (d \mu +\mu^2) \cdot B 
\end{eqnarray*}
Since $B$ a $\LG$-valued 2-form, we have $[B,B] = 0$ and we obtain :

\begin{eqnarray}
\boxed{ d\omega + \mu \cdot \omega = \nu \cdot B }
\end{eqnarray}

\subsection{Gauge transformations}

As in the case of bundles, we need to know when 
two gerbes define the same gauge invariant quantities and what the 
automorphisms of a given gerbe are. Our functorial approach suggests to define 
the gauge transformations on $\f$ as its equivalences of 2-functor.
Two gerbes $\f$ and $\f '$ are declared isomorphic \textit{iff.} there exists an 
invertible natural equivalence of 2-functors, $H:\f ' \to\f $, i.e. for 
each vertex $x$ a 1-arrow of $\TB$, $H_x : {\f '}_x \to \f_x$, 
and, for each edge $xy$, a 2-arrow 
$H_{xy} : H_x {\f '}_{xy} \to \f_{xy} H_y $, 
such that the following composition of 2-arrows be invertible

\vspace{1cm}
\hspace{4cm}
\pic{
\f_x & & 
\ar@{}[d]|{\rotatebox{90}{$\Longrightarrow$} \f_{[xyzx]}} & & \f_x \ar[llll]_{1_{\f_x}} 
\ar[dl] \\
& \f_y \ar@{}[ddl]|{\rotatebox{45}{$\Longrightarrow$} H_{xy}} \ar[ul] & & \f_z 
\ar[ll] \ar@{}[ddll]|{\rotatebox{45}{$\Longrightarrow$} H_{yz}} & 
\ar@{}[ddl]|{\rotatebox{45}{$\Longrightarrow$} H_{zx}} \\
& & & & \\
& {\f '}_y \ar[dl] \ar[uu]_{H_y} & \ar@{}[d]|{\rotatebox{90}{$\Longrightarrow$} {\f 
'}_{[xyzx]^*}}  
& {\f '}_z \ar[uu]^{H_z} \ar[ll] & \\
{\f '}_x \ar[uuuu]^{H_x} & & & & {\f '}_x \ar[llll]^{1_{{\f '}_x}} \ar[ul] 
\ar[uuuu]_{H_x}
}
\vspace{1cm}

\ni When $\f = \f '$, the gauge transformation $H$ induces an equivalence of $\F$ :
if, for each path $\g \in \O \, (\PX) $, $H_\g : \F_\g \to \F_\g $ denotes 
the action of $H$ on the category of sections of $\f_\g$, then $H$ is an equivalence of 
$\F$  
since for each edge $(\g , \g ')$ of $\PX$, there exists an invertible natural 
transformation from 
$H_{\g '} \F_{\g ' \g}$ to $\F_{\g ' \g} H_\g$ represented by the following diagram

\vspace{1cm}
\hspace{5cm}
\pic{
\F_\g \ar[dd]_{H_\g} \ar@{}[ddrr]|{\rotatebox{45}{$\iff $}} \ar[rr]^{\F_{\g ' \g}}
& & \F_{\g '} \ar[dd]^{H_{\g '}} \\
& & \\
\F_\g \ar[rr]_{\F_{\g ' \g}} & & \F_{\g '}
}
\vspace{1cm}

\ni In order to obtain the local expression of the gauge transformed 2-connection, 
let's pick $(b \xla{\alpha} a) \in \O \, (\F_{(xx)})$ and let's compare the action of 
$\F_{(xyzx , xx)} H_{(xx)}$ 

\begin{eqnarray*}
\F_{(xyzx , xx)} H_{(xx)} (b \xla{\alpha} a) &=& \big( H_x (b) , H_x (\alpha) \circ 
K_{[xyzx]}(H_x (a)) , \\
& & \f_{yz} \f_{zx} \beta_{xzyx} H_x (a) , \f_{yz} \f_{zx} \beta_{xzyx} H_x (1_a) , \\ 
& & \f_{zx} \beta_{xzyx} H_x (a) , \f_{zx} \beta_{xzyx} H_x (1_a) , \beta_{xzyx} H_x 
(a)  \big)
\end{eqnarray*}
with the action of $H_{(xyzx)} \F_{(xyzx , xx)}$ on the same initial section :

\begin{eqnarray*}
H_{(xyzx)} \F_{(xyzx , xx)} (b \xla{\alpha} a) &=& \big( H^{-1}_x (b) , H^{-1}_x 
(\alpha) \circ 
H^{-1}_x ( K_{[xyzx]} (a) ) \circ H^{-1}_{xy} (\f_{yz} \f_{zx} \beta_{xzyx} (a) ) ,  \\
& & H^{-1}_y \f_{yz} \f_{zx} \beta_{xzyx} (a) , H^{-1}_{yz} (\f_{zx} \beta_{xzyx} (a)) 
, \\
& & H^{-1}_z \f_{zx} \beta_{xzyx} (a) , H^{-1}_{zx} (\beta_{xzyx} (a) ) , H^{-1}_x 
\beta_{xzyx} (a) \big)
\end{eqnarray*}
Equating the compositions of these arrows, we obtain :

\begin{eqnarray*}
K_{[xyzx]} (H_x (a)) &=& H^{-1}_x ( K_{[xyzx]} (a) ) \circ H^{-1}_{xy} (\f_{yz} 
\f_{zx} \beta_{xzyx} (a) ) \\
& & \circ \, \f_{xy} H^{-1}_{yz} (\f_{zx} \beta_{xzyx} (a)) \circ \f_{xy} \f_{yz}  
H^{-1}_{zx} (\beta_{xzyx} (a) )
\end{eqnarray*}
In a local chart, we have 

\begin{eqnarray*}
K_{[xyzx]} (a) &=& 1_G + B_{xyzx} \\
K_{[xyzx]} (H_x (a)) &=& 1_G + {B'}_{xyzx} \\
\f_{xy} &=& 1_{\AG} + \mu_{xy} \\
{\f '}_{xy} &=& 1_{\AG} + {\mu '}_{xy} \\
H_x &=& 1_{\AG} + \xi_x \\
H_{xy} &=& 1_G + \eta_{xy}
\end{eqnarray*}
where $\xi_x \in \LAG$ and $\eta_{xy} \in \LG$. The expansion of the previous identity 
up to first order in $(\xi , \eta )$ provides us with the expression of the 
infinitesimal gauge transformations :

\begin{eqnarray*}
1 + {B'}_{xyzx} &=& \big[ (1 - \xi_x) \cdot (1 + B_{xyzx}) \big] \big[1 - \eta_{xy} 
\big] \\
&& \quad \big[ (1 + \mu_{xy}) \cdot (1 - \eta_{yz}) \big] 
\big[(1+\mu_{xy})(1+\mu_{yz})\cdot (1 - \eta_{zx}) \big] \\
{B'}_{xyzx} &=& B_{xyzx} - \xi_x \cdot B_{xyzx} - 
(\eta_{xy} + \eta_{yz} + \eta_{zx} + \mu_{xy} \cdot \eta_{yz} - \mu_{xy} \cdot 
\eta_{zx} - \mu_{yz} \cdot \eta_{zx} )
\end{eqnarray*}
Using the differential forms $\xi = \sum_x x \xi_x$ and $\eta = \sum_{(xy)} (xy) 
\eta_{xy}$ :

\begin{eqnarray}
\boxed{ B' = B - \xi \cdot B - (d \eta + \mu \cdot \eta) }
\end{eqnarray}
The infinitesimal effect of $H$ on the 1-connection is obtained by expanding 
up to first order in $(\xi , \eta )$ the equivalence between functors :

\begin{eqnarray*}
{\f '}_{xy} & \simeq & H^{-1}_x \f_{xy} H_y \\
1 + {\mu '}_{xy} &=& (1 - \xi_x) (1 + \mu_{xy}) (1 + \xi_y) \\
{\mu '}_{xy} &=& \mu_{xy} + \xi_y - \xi_x + \mu_{xy} \xi_y - \xi_x \mu_{xy}
\end{eqnarray*}
In terms of differential forms :

\begin{eqnarray}
\boxed{\mu ' = \mu + d \xi + [\mu , \xi] }
\end{eqnarray}
Introducing the notation $\nabla = d + \mu$, we have

\begin{eqnarray*}
\nabla^2 &=& d \mu + \mu^2 \\
\nabla \xi &=& d \xi + [\mu , \xi ] \\ 
\nabla B &=& dB + \mu \cdot B = \omega \\
\nabla \eta &=& d \eta + \mu \cdot \eta
\end{eqnarray*} 
and similarly with $\nabla '$ and $\mu '$.
The action of $H$ on the 2-curvature $\nu$ is given by 

\begin{eqnarray*}
\nu ' &=& d\mu ' + (\mu ')^2 + [B', \cdots]\\
&=& d(\mu + \nabla \xi) + (\mu + \nabla \xi )^2 + [B - \xi \cdot B - \nabla \eta , 
\cdots ] \\
&=& \big( d \mu + \mu^2 + [B , \cdots] \big)+ d \nabla \xi + \mu \, \nabla \xi + 
(\nabla \xi ) \, \mu - [\xi \cdot B + \nabla \eta , \cdots ]
\end{eqnarray*}

\begin{eqnarray}
\boxed{\nu ' = \nu + \nabla^2 \xi - [\xi \cdot B + \nabla \eta , \cdots ] }
\end{eqnarray}
Similarly, the first order gauge variation of the curvature 3-form $\omega$ is given by

\begin{eqnarray*}
\omega ' &=& dB' +\mu ' \cdot B' \\
&=& dB - d (\xi \cdot B) - d (\nabla \eta ) + 
(\mu + \nabla \xi ) \cdot (B - \xi \cdot B - \nabla \eta) \\
&=& \omega - \nabla (\xi \cdot B) + (\nabla \xi ) \cdot B - \nabla^2 \eta 
\end{eqnarray*}

\begin{eqnarray}
\boxed{ \omega ' = \omega - \xi \cdot \omega - \nabla^2 \eta}
\end{eqnarray}

\section{Perspectives}

In the long search of a string model of the confined phase of Yang-Mills
theory \cite{Pol}, A. Polyakov emphasized a crucial difference between the 
metric strings, which are coupled to the background metric $G_{\mu \nu}$ and
include gravitons in their spectrum, and the gauge strings, which don't. The 
gauge symmetry of the theory formulated in terms of connection and 
curvature is adapted to the short distance physics. 
In the string picture, the gauge symmetry translates into the 
zizgzag symmetry : if a thin homotopy relates two surfaces with the same fixed 
boundary, then their actions are equal. This symmetry is precisely what is needed 
to project the zero mass spectrum of the string on the space generated by the 
vector vertex operators, thus eliminating the graviton.
In our setting, the zigzag symmetry acts in the fibered category $\F$. 
An elementary backtracking is a pair of paths, say 
$\g = (\cdots , x , y , \cdots)$ and $\g' = (\cdots , x , z , x , y , \cdots)$, 
and the corresponding functor $\F_{\g \g'}$ acts as

\begin{eqnarray*}
& & \F_{\g' \g} (\cdots , a_x , u_{xz} , a_z , u_{zx} , {a'}_x , \cdots) \\
& & = \quad (\cdots , a_x , u_{xz} \circ \f_{xz} (u_{zx}) \circ K_{[xzzx]^*} ({a'}_x)  
, 
\beta_{xzzx}({a'}_x) , \beta_{xzzx}(\cdots)) 
\end{eqnarray*}
Therefore, the requirement of the zigzag symmetry is equivalent to 

$$
K_{[xzzx]} = K_{[xzzx]^*} = {\bar{K}}_{[xzzx]} = {\bar{K}}_{[xzzx]^*} = 1_{1_{\f_x}} 
\simeq 1_G 
$$
As noted in (\S 4.1), this condition, which bears on $\beta$ and $C$, amounts to 
specifying a quasi-inverse of the connective structure : $\f^{-1}_{xy} = \f_{yx}$.
The action of the pure $BF$-theory is defined on a four-manifold $M$ by

$$
S (\mu , B) = \int_{M} \mathrm{tr}_{\mathrm{ad}} (\nu \, [B, \cdots]) 
$$
We deduce from the relations (9-12) its gauge variation :

\begin{eqnarray*}
S(\mu ' , B') - S (\mu , B ) &=& - \int_{M}  {\mathrm{tr}}_{\mathrm{ad}} 
\big( \nabla (\xi \cdot \omega ) + \nu \cdot \nabla \eta \big) \\
&=& - \int_{\partial M} {\mathrm{tr}}_{\mathrm{ad}} 
\big( \xi \cdot \omega + \frac{1}{2} \, \mu \cdot \nabla \eta \big) 
\end{eqnarray*}
The identity (8) is a higher dimensional analogue of the Bianchi identity (2), and 
shows clearly the origin of the topological invariance of $S$. \nl

\textbf{Remarks : }

$i$) Since our approach does not suppose the existence of a globally defined 2-form $B$, 
all integral formulas depend on the choice of a $\check{\mathrm{C}}$ech 1-cocycle 
taking its values in $\AG$ and adapted to an open hypercover of $M$.

$ii$) The section $(a,u) \in \O (\F_\g)$, transported by the successive sweeping functors 
$\F_{\g \g'}$, provides a $G$-bundle with connection over the moving path and plays the same role as a 
fermion transported by a gauge field in Yang-Mills theory. \nl

As suggested in (\S 4.4), the geometry of gerbes is already adapted to the kinematics 
of topological field theories of type $BF$. Indeed, if we compare the gauge 
transformations given in (\S 4.3) with those of the pure $BF$-theory \cite{BF}, 
we remark that we can make the following correspondence :

\begin{eqnarray*}
\mathrm{gerbes} &\lra & BF-\mathrm{theory} \\
& & \\
\mu & \lra & [A , \cdots ] \\
\nu & \lra & [F_A , \cdots ] - [B ,  \cdots ] \\
\end{eqnarray*}
This leads us to believe that the right framework for $BF$-type field theories 
is the space of gerbes and not the narrower theory of (principal or associated) 
bundles.

\vspace{1cm}

\textbf{Conclusion :} We have developped a simplicial approach to non-Abelian gerbes 
and proved, using a 2-groupoid combinatorics, that it leads to the same cohomology 
classes as the differential approach. The specificity of our viewpoint resides in the 
usage of the space of iterated edge-paths and of homotopy $n$-groupoids. The importance of the 
space of paths is suggested by gauge field theories where physical quantities are 
naturally associated to loops. Conversely, the developement of the theory of gerbes may 
open the door to the resolution of long standing problems of physics like the 
mathematical proof of confinement of quarks in QCD or the construction of integrable 
models in dimension higher than two \cite{AFSG}. We hope to develop these ideas in a 
near future. \newline

\textbf{Acknowledgements :} I wish to thank Lawrence Breen for useful discussions at 
the beginning of this work, Daniel Bennequin and Olivier Babelon for their 
encouragements, and Marc Bellon for a careful reading of the manuscript.

\end{document}